# A Protein Structure Prediction Approach Leveraging Transformer and CNN Integration

Yanlin Zhou[1],*
Computer Science
Johns Hopkins University
MD,USA,
popojoyzhou@gmail.com

Kai Tan[1]
Electrical & Computer Engineering
University of Washington,
seattle, WA
tank5@uw.edu

Xinyu Shen[2]
Biostatistics
Columbia University
Frisco,TX,USA
xshe1007@gmail.com

Zheng He[3]
Applied Analytics
Columbia University
NY,USA
zh2541@columbia.edu

Haotian Zheng[4]
Electrical & Computer Engineering
New York University
New York, NY, USA
hz2687@nyu.edu

*Abstract*—Proteins are essential for life, and their structure determines their function. The protein secondary structure is formed by the folding of the protein primary structure, and the protein tertiary structure is formed by the bending and folding of the secondary structure. Therefore, the study of protein secondary structure is very helpful to the overall understanding of protein structure. Although the accuracy of protein secondary structure prediction has continuously improved with the development of machine learning and deep learning, progress in the field of protein structure prediction, unfortunately, remains insufficient to meet the large demand for protein information. Therefore, based on the advantages of deep learning-based methods in feature extraction and learning ability, this paper adopts a two-dimensional fusion deep neural network model, DstruCCN, which uses Convolutional Neural Networks (CCN) and a supervised Transformer protein language model for single-sequence protein structure prediction. The training features of the two are combined to predict the protein Transformer binding site matrix, and then the three-dimensional structure is reconstructed using energy minimization.

*Keywords-Protein structure prediction; Convolutional neural network; Sequence information; RNA site*

## I. Introduction (Heading 1)

The mechanism of the interaction between proteins and RNA is the basis for understanding various biological activities and designing new drugs. It is critical to adequately characterize the determinants that influence protein-RNA binding specificity at the molecular level. With the rapid progress of sequencing technology, the data on the interaction between proteins and RNA are increasing, providing the possibility to predict the binding sites of proteins and RNA on a large scale by computational methods. The predictive analysis of the binding site between protein and RNA can not only deepen the understanding of the mechanism of action between protein and RNA but also apply the predictive model to more biological processes. Many studies in computational biology have focused on the prediction of protein and RNA binding sites, including traditional machine learning-based prediction methods and deep learning-based prediction methods. Methods based on traditional machine learning need to manually design data features and rely on domain knowledge, which is difficult to implement. Although the method based on deep learning has advantages in feature extraction and learning ability, the research using this method mostly focuses on the prediction of sequence pair interaction or the prediction of binding sites in single-sequence fragments, and cannot predict whether specific amino acid-nucleotide pairs in proteins and RNA are bound.

At the same time, most of the existing research methods extract the features of protein and RNA at the sequence level and rarely consider their spatial structure characteristics. Based on the problems existing in the existing research, this paper proposes a single sequence protein structure prediction algorithm, Transformer be, based on data source and network model. Specifically, Transformer be integrates sequence embedding from a supervised Transformer protein language model into a multi-scale network enhanced by knowledge distillation, predicts the two-dimensional geometry between residues, and then reconstructs the three-dimensional structure using energy minimization.

---

1 * Corresponding author: [Yanlin Zhou]. Email: [popojoyzhou@gmail.com]

## II. Related work

### A. Protein Engineering Overview

Protein engineering is the process of creating proteins with specific functions, such as improving organism characteristics, enhancing enzyme catalytic performance, and increasing antibody potency. This field has had a significant impact on drug discovery, enzyme development, biosensors, diagnostics, and other biotechnologies. It also provides fundamental principles for understanding the relationship between protein structure and function. In addition, protein engineering has a positive impact on sustainability and environmental protection. For instance, designing and optimizing industrial enzymes can lead to more environmentally friendly chemical reaction processes and less hazardous waste generation. The field of protein engineering is expected to drive innovation and provide possibilities for future improvements in life.

Two traditional methods are mainly used in this field: directed evolution and rational design. Directed evolution is a process used to create proteins or enzymes with improved or new functions. The directed evolution approach entails introducing mutations into the genetic code of a target protein and screening the resulting variants to enhance their function. This process is guided by desired outcomes such as improved activity, stability, specificity, binding affinity, and adaptability. In contrast, rational design employs knowledge of protein structure and function to make specific modifications to a protein sequence or structure.

### B. Machine learning AIDS protein engineering

In recent years, machine learning has provided new solutions for directed evolution and protein engineering. Machine learning assisted protein engineering refers to the application of machine learning models and techniques to improve the efficiency and effectiveness of protein engineering. This approach reduces costs and accelerates progress in protein engineering. It optimizes protein screening and variant selection, increasing work efficiency and productivity. Researchers can quickly generate and test a large number of variants to build a protein's fitness map (i.e. fitness landscape) by analyzing and predicting the effects of mutations on protein function through machine learning, and then collect experimental data. The use of data-driven machine learning has significantly expedited the process of protein engineering.

This process typically involves several elements, including data acquisition and preprocessing, model design, feature extraction and selection, algorithm selection and design, model training and validation, experimental validation, and iterative model optimization. Advances in electrochemical biosensors and microfluidic technologies have played a significant role in high-throughput sequencing and screening techniques, resulting in the accumulation of a large number of experimental datasets on protein sequence, structure, and function. These datasets, along with deep mutation scanning libraries dedicated to protein engineering, provide a valuable resource for training and validating machine learning models.

### C. The importance of protein property prediction

Proteins are crucial for life processes, exhibiting a diverse and complex range of properties that are critical to biological and medical research. According to Koehler Leman et al.'s research, protein properties include amino acid sequence, three-dimensional spatial structure, biological function, and interactions with other molecules. The three-dimensional structure and function of a protein are crucial to its biological role. The fine-tuned structure of a protein is determined by its amino acid sequence, which is the key to its diverse molecular functions. Therefore, understanding the relationship between amino acid sequence and protein structure is significant for promoting biology comprehension and medical applications. The three-dimensional structure of a protein is a crucial factor in its function. Proteins interact with other molecules through their unique spatial configuration to perform various biological functions, such as catalytic biochemical reactions and signal transmission. Accurately predicting protein structures is essential for comprehending their functional mechanisms. In recent years, the analysis time from sequence to structure in protein structure prediction has significantly decreased, and the prediction accuracy has improved due to the continuous development of computing technology, especially AI. Kuhlman et al. suggest that these technological advances can drive cutting-edge research in protein structure prediction and design.

Secondly, predicting protein function is crucial in biomedical research. Proteins play key roles in life processes through their functions in living organisms, such as the catalytic activity of enzymes and the signalling activity of receptors. Although high-throughput sequencing technology allows for the rapid acquisition of a large number of protein sequences, their functions are still not fully understood. Jeffery et al. state that the development of computational methods to predict protein function is crucial for drug development and disease mechanism studies.

Protein-protein interactions (PPIs) are a fundamental component of many biological processes within the cell and have significant effects on signal transduction and metabolic pathways. Although traditional experimental methods can provide data on protein-protein interactions (PPIs), they are often time-consuming and prone to producing false positive results. Durham et al. 7 suggest that computational methods are increasingly important in predicting PPIs, with the goal of more efficiently identifying and verifying these interactions to advance research and applications in biology and medicine.

### D. Application of AI technology in protein prediction

From an AI technology perspective, AlphaFold utilises deep learning to understand the intricate relationship between a protein's amino acid sequence and its three-dimensional structure. AlphaFold's innovation is in its combination of multiple sequence alignment (MSA) data with physical and biological information to predict the distance and angle of amino acid sequences. AlphaFold2's primary innovation is in its deep learning architecture. The self-attentional Transformer architecture and the 'Evoformer' module are utilized to efficiently integrate protein sequence and structure information, improving prediction accuracy. This architecture is particularly effective at capturing patterns in protein sequences and combining evolutionary information to predict the three-dimensional structure of proteins. On the other hand, RoseTTAFold utilises a three-track neural network that considers the pattern of protein sequence, amino acid interactions, and the potential three-dimensional structure of proteins. Its multi-track neural network architecture can process information of varying dimensions simultaneously, enabling effective learning and prediction of protein structure. The models above not only demonstrate the significant potential of deep learning in biology but also offer new possibilities for biomedical research and drug development.

## III. CONVOLUTIONAL NETWORK PREDICTION

### A. The importance of protein structure prediction

Protein structure prediction is a long-standing and fascinating problem in the life sciences, but it is also known for its difficulty, high cost, and limited progress. There are three main traditional methods to observe protein structures, namely nuclear magnetic resonance, X-ray, cryo-electron microscopy, but these methods often rely on a lot of trial and error and expensive equipment, and each structure can take years to study. The latest AI application to protein structure prediction, AlphaFold2, can predict, in days or even minutes, protein structures with high confidence that previously took decades to obtain. Therefore, protein structure prediction through artificial intelligence will greatly improve people's understanding of life processes. For example, geneticists may accumulate vast amounts of data, but without knowing the structure of a protein, they cannot study the effect of a mutation on its function. Now, with AlphaFold2's structural prediction, it is possible to see where each mutation in a human genetic disease is located in the relevant protein structure, making it possible to infer how the protein function is affected.

### B. Convolutional networks and protein prediction

The Convolutional Neural Network (CNN) is one of the most widely used deep learning techniques. It is a kind of deep neural network with a feature extractor (composed of a convolutional layer and a mixed pool layer), which is popular in the field of computer vision. Computer vision is very intuitive and easy to understand, and the field of machine learning has been interested in it for a long time.

In order to introduce the application of convolutional neural networks in the field of biological genomics, this paper attempts to describe the CNN method for predicting DNA-protein binding sites in a way that is easily understood by machine learning or genomics researchers. First, we think of a genome sequence window as a picture. Instead of a two-dimensional picture composed of three color channels (R, G, B) pixels, we treat the genome sequence as a fixed-length one-dimensional sequence window composed of four channels (A, C, G, T). Therefore, the problem of predicting DNA protein binding sites is similar to the binary classification problem in the picture. One of the biggest advantages of CNN for genomics research is that it can detect whether a motif (a protein molecule with a specific function or a similar secondary structure aggregation as a part of an independent domain) is in the specified sequence window. This detection ability is very conducive to the identification of motifs, and then helps to the classification of binding sites.

### C. Protein sequence training model resolution structure

Compared to the previous generation model ESM-1b, ESM-2 introduces improvements in architecture, training parameters, and increased computational resources and data. The resulting ESM-2 model family significantly outperforms the previous state-of-the-art ESM-1b (about 650 million parameter models) on a comparable number of parameters, and it also outperforms other recent protein language models on structural prediction benchmarks.

The ESM-2 language model is trained using a mask language modeling target that predicts the identity of randomly selected amino acids in a protein sequence by looking at the context of the rest of the sequence. This allows the model to learn the dependencies between amino acids. Although the training goal itself is simple and unsupervised, to accomplish this task on millions of evolutionarily different protein sequences requires the model to internalize sequence patterns throughout evolution.

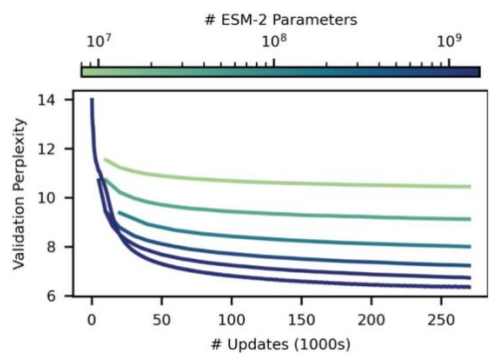

Figure 1. ESM-2 mask language modeling training curve

Since ESM-2 is trained only on sequences, any information about the development of the structure must be the result of representing the pattern of the sequence. Transformer models trained using mask language modeling are known to develop attentional patterns that correspond to residue contact diagrams of proteins.

### D. Introduction to Transformer

The Transformer neural network has revolutionized structural biology by predicting protein structures with unprecedented accuracy. The landmark event in this regard was in 2020, when DeepMind's AlphaFold2 method made a major scientific breakthrough in predicting the three-dimensional structure of proteins from amino acid sequences. At the heart of the AlphaFold2 framework is a Transformer neural network driven by an attention mechanism. What makes the Transformer architecture so powerful is its ability to simulate long-term relationships in input sequences beyond their sequential neighborhood.

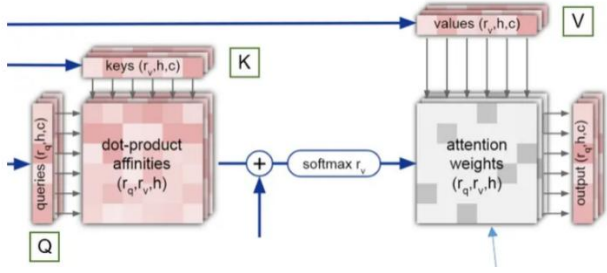

Figure 2. AlphaFold2 performs protein structure prediction models

AlphaFold2 uses the target amino acid sequence, MSA and template as inputs to directly predict the 3D structure of the target from end to end, and uses Transformer for pre-training.

1. MSA refers to Multiple Sequence Alignment, which refers to sequence comparison after alignment of multiple homologous amino acid sequences. For any two amino acid sequences, the position can be moved left and right by filling the vacancy, etc.

2. Evoformer block, the two sets of information after comparison will form a 48block Evoformer block, and then obtain relatively similar alignment sequences. The key innovation of Evoformer module is the new mechanism for exchanging information within MSA and the paired representation that allows direct inference of spatial and evolutionary relationships. The rotating and shifting form of each residue of a protein (global rigid body framework).

Traditional CNNs and RNNs are abandoned in Transformer, and the entire network structure is completely composed of the attention mechanism. More precisely, Transformer consists of and only consists of self-attention and Feed Forward Neural networks. A trainable neural network based on Transformer can be built by stacking Transformers. In the experiment, a total of 12 Encoder-Decoder layers with 6 encoders and decoders were built, and a new high BLEU value was achieved in machine translation.

In summary, Transformer be, a single sequence protein structure prediction algorithm based on a neural network structure, was proposed in this study. Specifically, Transformer be integrates sequence embedding from the supervised Transformer protein language model into a multi-scale network enhanced by knowledge distillation to predict the two-dimensional geometry between residues.

## IV. METHODOLOGY

Transformer is used to predict residue contact because the attention force generated within the Transformer model naturally corresponds to the information between the individual residues in the sequence. Formally, both the attention diagram and the residue contact diagram can be represented as an L×L square matrix.

### A. Modeling language modeling MLM

When predicting protein structure, ESM-1b uses multiple protein language models to predict possible mutations in the target antibody, and uses the prediction scheme unanimously recommended by multiple models to obtain the required amino acid replacement strategy. You need to take a given original sequence x as input:

$$x = (x_1, \cdots, x_n) \in x^N \quad (1)$$

Where x is a sequence of amino acids and N represents the length of that sequence. In addition, we also need a set of pre-trained masked language models to generate conditional likelihood probability. To guide the evolution based on a specific language model, we first compute the set of alternatives for which the language model is more likely than the wild type, and the formula can be expressed as follows:

$$\mathcal{M}(p_j) = \{i \in [N], \; x'_i \in x: \; \frac{p_j(x'_i|\mathbf{x})}{p_j(x'_i|\mathbf{x})} > \alpha\}, \quad (2)$$

Where Pj represents one of the language models, "represents an amino acid residue of a wild-type antibody, and" = 1. In order to further filter out only the alternatives with the highest likelihood, we select the prediction scheme consistently recommended by multiple models as the final alternative where, for the new amino acids, we calculate:

$$f(x'_i) = \sum_{j \in [M]} 1\{(i, x'_i) \text{ is in } \mathcal{M}(p_j)\} \quad (3)$$

We then obtain a higher probability replacement set than the wild type in multiple language models, which can be expressed as follows:

$$\mathcal{A} = \{i \in [N], x'_i \in x: \; f(x'_i) \geq k\} \quad (4)$$

Where is the cut-off value, which controls the number of corresponding variants to be measured.

### B. Model building

Transformer first uses mask language modeling (MLM) to pre-train the sequences from the large database (Uniref50). Once the training is complete, the attention map can be extracted. Two operations, Symmetrization and Average Product Correction (APC), convert the attention map to the required form, and then perform the regression task. Logistic regression with power 1 regularization is applied independently on each amino acid pair (i,j).

The ESM-1b framework is shown as follows:

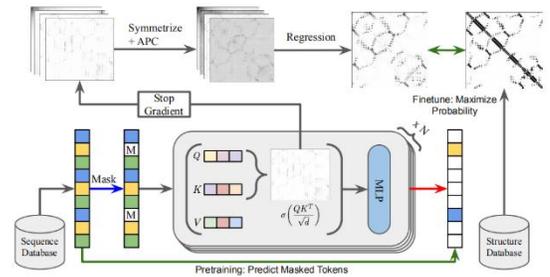

Figure 3. Framework model of ESM-1b

Regression is trained on a small number of proteins (n < 20) to determine which Attention heads contain a wealth of information. During training, we only trained the weights of ogistic regression, and did not backpropagate the whole model, that is, on the branch of logistic regression, the Gradient value backpropagated to the attention diagram, and the gradient backpropagation was stopped at the "Stop Gradient" section in the figure above.

*C. Train the Gremlin model*

For a set of training sequences X, Gremlin optimizes the following pseudo-likelihood loss:

$$L_{PLL}(X;\theta) = E_{x \sim X} \sum_{i=1}^{L} \log p(x_i | x_{j \neq i}; \theta) \quad (5)$$

Where individual locations are masked and their true amino acid residues are predicted from their context, the context is used to predict the center word, essentially the same as the continuous word bag model CBOW in Word2Vec. Since the inputs to Gremlin are aligned, they have a uniform length L. Gremlin takes MSA as input. The quality of the output prediction depends heavily on the construction of the MSA.

*D. Forecast result*

This experiment evaluated a pre-trained protein language model in a trRosetta training dataset with 15,051 proteins, in which 43 proteins with sequence length greater than 1024 were removed from the trRosetta dataset because the training context size of ESM-1b was set to 1024. Of these sequences, when we tried to construct the MSAs using the ESM training set, Jackhmer failed in 126 proteins, that is, the construction failed, so we ended up with only 14,882 protein sequences. At the end, the authors reserved 20 sequences for training, 20 sequences for validation, and the remaining 14,842 sequences for testing.

TABLE I. AVERAGE PRECISION ON 14842 TEST STRUCTURES FOR TRANSFORMER MODELS TRAINED ON 20 STRUCTURES.

| Model | 6 ≤ sep < 12 | | | 12 ≤ sep < 24 | | | 24 ≤ sep | | |
|---|---|---|---|---|---|---|---|---|---|
| | L | L/2 | L/5 | L | L/2 | L/5 | L | L/2 | L/5 |
| Gremlin (ESM Data) | 15.2 | 23.0 | 37.8 | 18.1 | 27.9 | 44.3 | 31.3 | 43.1 | 55.5 |
| Gremlin (trRosetta Data) | 17.2 | 26.7 | 44.4 | 21.1 | 33.3 | 52.3 | 39.3 | 52.2 | 62.8 |
| TAPE | 9.9 | 12.3 | 16.4 | 10.0 | 12.6 | 16.6 | 11.2 | 14.0 | 17.9 |
| ProtBERT-BFD | 20.4 | 30.7 | 48.4 | 24.3 | 35.5 | 52.0 | 34.1 | 45.0 | 57.4 |
| ESM-1 (6 layer) | 11.0 | 13.2 | 15.9 | 11.5 | 14.6 | 19.0 | 13.2 | 16.7 | 21.5 |
| ESM-1 (12 layer) | 15.2 | 21.1 | 30.5 | 18.1 | 24.7 | 34.0 | 23.7 | 30.5 | 39.3 |
| ESM-1 (34 layer) | 20.3 | 30.2 | 46.0 | 23.8 | 34.3 | 49.2 | 34.7 | 44.6 | 56.0 |
| ESM-1b | **21.6** | **33.2** | **52.7** | **26.2** | **38.6** | **56.4** | **41.1** | **53.3** | **66.1** |

In Table 1, all Transformer model contact predictors are trained on 20 proteins by Logistic regression after pre-training. Although the ESM-1b model was trained on only 20 protein sequences, it has higher accuracy than Gremlin in contact prediction at short, medium and long range.

In this paper, protein language models trained on Transformer with unsupervised targets learn information about the tertiary structure of protein sequences in their attention graphs. Sparse (L1 regularization) Logistic regression can be used to extract useful information about residue contacts from the attention diagram. In addition, it has been found that different attentional heads are specifically responsible for different types of contact. Analysis of the restricted supervision part confirmed that the information needed for contact prediction was learned in the unsupervised phase, and Logistic regression only needed to extract the part of the model representing contact information.

## V. CONCLUSION

This paper aims to enhance the clarity of protein structure prediction through the utilization of artificial intelligence algorithms that leverage Transformer and CNNs. By predicting both two-dimensional and three-dimensional protein structures, our algorithm has demonstrated remarkable achievements in improving prediction accuracy and efficiency. The integration of Transformer and CNNs allows us to harness their respective strengths in feature extraction and learning, ultimately elevating the accuracy of protein structure prediction.

Furthermore, we introduce the ESM-1be algorithm, which incorporates sequence embeddings generated by a supervised Transformer-based protein language model. This innovative approach predicts two-dimensional geometric relationships between protein residues through a multi-scale network enhanced by knowledge distillation. Emphasizing the importance of the Attention mechanism, we analyze its pivotal role within the Transformer model. It effectively simulates long-term relationships within input sequences, thereby providing unprecedented accuracy in protein structure prediction.

In summary, artificial intelligence algorithms like Transformer have achieved significant breakthroughs in the field of protein structure prediction, substantially reducing the time required for structure determination. This accomplishment not only holds profound implications for our understanding of biological processes but also opens up new avenues for investigating the effects of genetic mutations on protein functionality. The success of AI in protein structure prediction underscores the continuous evolution of algorithms, promising even more precise and efficient prediction models in the future. These advancements will significantly contribute to our insights into protein biological functions, providing a more reliable foundation for areas such as novel drug design.